\documentclass[%
aps,
 amsmath,amssymb,
 reprint,%
]{revtex4-1}

\usepackage{graphicx}
\usepackage{dcolumn}
\usepackage{bm}
\usepackage{SIunits}
\usepackage{color}

\newcommand{\op}[2]{\mbox{$\hat{#1}_{#2}$}}
\newcommand{\aopd}{\hat{a}^\dagger}
\newcommand{\aop}{\hat{a}}

\newcommand{\bopd}{\hat{b}^\dagger}
\newcommand{\bop}{\hat{b}}

\begin{document}


\title[Engineering phonon leakage in nanomechanical resonators]{Engineering phonon leakage in nanomechanical resonators}

\author{Rishi N. Patel}
\email{rishipat@stanford.edu}
\author{Christopher J. Sarabalis}%
\author{Wentao Jiang}%
\author{Jeff T. Hill}
\author{Amir H. Safavi-Naeini}%
\email{safavi@stanford.edu}
\affiliation{%
Department of Applied Physics and Ginzton Laboratory, Stanford University
}%

\date{\today}
\begin{abstract}
We propose and experimentally demonstrate a technique for coupling phonons out of an optomechanical crystal cavity. By designing a perturbation that breaks a symmetry in the elastic structure, we selectively induce phonon leakage without affecting the optical properties. It is shown experimentally via cryogenic measurements that the proposed cavity perturbation causes loss of phonons into  mechanical waves on the surface of silicon, while leaving photon lifetimes unaffected. This demonstrates that phonon leakage can be engineered in on-chip optomechanical systems. We experimentally observe large fluctuations in leakage rates that we attribute to fabrication disorder and verify this using simulations. Our technique opens the way to engineering more complex on-chip phonon networks utilizing guided mechanical waves to connect quantum systems.
\end{abstract}

\pacs{Valid PACS appear here}
\maketitle

\section{Introduction}

Laser light can be used to read out and drive motion in small mechanical resonators fabricated on the surface of a chip. One paradigm for implementing such devices is that of optomechanical crystals -- periodic structures etched into thin films of dielectric, where the periodicity leads to confinement and co-localization of the optical and mechanical energy~\cite{Eichenfield2009,Safavi-Naeini2014a}. Strong co-localization leads to very large interaction strengths making these devices particularly attractive in the field of quantum optomechanics. Such devices have been implemented in silicon~\cite{Chan2011,Chan2012,Safavi-Naeini2014snowflake}, silicon nitride~\cite{Davanco2014}, gallium arsenide~\cite{Balram2015c}, aluminum nitride~\cite{Bochmann2013a,Vainsencher2016}, and diamond~\cite{Burek}, and are actively pursued in a variety of experiments on quantum optomechanics~\cite{Chan2011,Safavi-Naeini2012,Meenehan2015,Cohen2015,Riedinger2016}. In addition to novel optomechanical dynamics, these systems afford quantum optical control over the localized mechanical modes on a chip. This makes them particularly suited as transduction elements in phonon networks -- networks of mechanically coupled systems where quantum information is distributed through phonon waveguides~\cite{Safavi-Naeini2010a,Habraken2012,Vermersch2017,Fang2016,Fang_2016b}. 

Realizing on-chip phonon networks requires engineering the phonon modes and the coupling of these modes to each other in a controllable way. To connect the optically coupled phononic resonances to other mechanical and mechanically-coupled degrees of freedom on a chip, it is important to find ways to controllably engineer phonon loading, i.e., the \emph{leakage} of phonons out of an optomechanical crystal cavity, without negatively impacting the  optical and mechanical properties of the device. This is   challenging since the confinement of phonons in optomechanical crystal cavities is brought about by Bragg reflection, the same mechanism that confines the optical field. In addition, the optical quality factor is highly sensitive to the local patterning of the dielectric and generally, small perturbations in the hole dimension and position cause leakage into radiation modes and reduction in the optical quality factor.

Here we theoretically and experimentally demonstrate that by generating a particular perturbation in the position of holes in an optomechanical crystal cavity, the mechanical quality factor can be tuned over nearly two orders of magnitude without a noticeable effect on the optical properties of the system. We begin in section~\ref{s:design} by introducing the optomechanical crystal system used in this work and providing arguments of why certain perturbations can lead to phonon leakage. We provide results of simulations to back up a heuristic understanding of the design procedure. In section~\ref{s:experiment} we present experimental results validating the design concept. Finally, we see that fabrication disorder leads to an unexpected reduction in mechanical loss rates, which we analyze in section~\ref{s:discussion}.

\section{Design and Simulations}\label{s:design}

\begin{figure}
\includegraphics[width=0.5\textwidth]{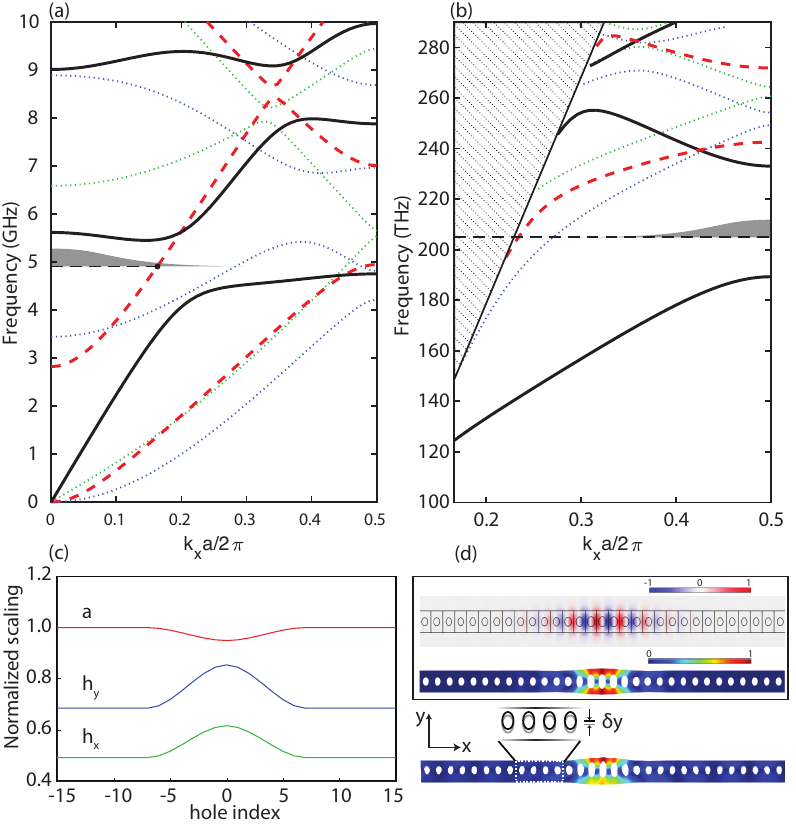}
\caption{\label{bands_concept} Concept and 1D cavity design. (a) Phononic band structure of the mirror (nominal) unit cell. The solid black lines denote the fully symmetric band, from which a $\Gamma$-point defect state is pulled, while the dashed red line shows the band of opposite $y$-symmetry. Via the offset perturbation described in the text, the dashed black line shows how a mode at the defect frequency can couple to this band. The blue and green bands have opposite $z$-symmetry forbidding transitions. (b) Photonic band structure of the mirror unit cell. The solid black line denotes the fundamental TE band from which the $X$-point defect mode is generated. As in (a), the red dashed line shows the band of opposite y symmetry. The top left region represents the continuum of radiation modes above the light line. (c) The variations of $a$, the lattice constant, $h_x$, the hole diameter in the $x$-direction, and $h_y$,the hole diameter in the $y$ direction, all normalized by the nominal lattice constant in the mirror region are plotted as a function of hole index. (d) Electric field $E_{y}$ of the fundamental TE optical mode (top) and fundamental mechanical breathing mode of the cavity (middle). The breathing mode of the structure with a $40~\textrm{nm}$ offset perturbation is shown on the bottom, along with an exaggerated view of a unit cell defining the perturbation.}
\end{figure}

Our system consists of a one-dimensional photonic crystal with simultaneous photonic and phononic bandgaps -- a so-called optomechanical crystal~\cite{Eichenfield2009,Safavi-Naeini2014a}. The system is engineered with a fundamental TE optical mode in the telecom band ($\approx 1550~\nano\meter$), and a fundamental mechanical breathing mode at around 5 GHz. The band structure for phonons and photons propagating in the nominal unit cell of our structure is shown in Fig.~\ref{bands_concept}a and b. In the phononic band diagram (Fig. \ref{bands_concept}a), we see a symmetry-dependent bandgap between 4.75 GHz and 5.45 GHz. For the photons (Fig. \ref{bands_concept}b), there is also a symmetry-dependent bandgap for guided modes confined to the beam between 189 THz and 233 THz corresponding to optical free space wavelengths of $1288~\textrm{nm}$ to $1587~\textrm{nm}$. A defect is generated by locally and smoothly varying the lattice spacing and ellipse dimensions. The exact form of this variation is obtained by an optimization algorithm similar to the one described in Ref.~\cite{Chan2012}. The variations of the parameters as a function of the hole number are plotted in Fig. \ref{bands_concept}c. where the nominal lattice spacing is $a=336~\nano\meter$. This variation preserves the symmetry of the system for reflections about mirror planes perpendicular the $y$ and $z$ axes. By preserving these symmetries, we ensure that the confined phonons cannot leak into guided waves propagating in the mirror regions. Separately, the slowly varying unit cell properties as determined by the numerical optimization procedure means that photons do not efficiently couple to out-going radiation -- the mode has a vanishing amount of energy in momentum space above the light line (hatched region) where radiation can lead to leakage of optical energy. 

To generate phonon leakage, we induce a perturbation that couples the confined phonon mode to guided waves with a differing symmetry. Here, the key idea is to introduce an offset perturbation, whereby each unit cell hole is offset in the $y$ direction by an amount $\delta{y}$. This perturbation breaks the reflection symmetry of the geometry about the $y=0$ mirror plane, and allows coupling between the $(+_{y},+_{z})$ band (black) and the $(-_{y},+_{z})$ band (dotted red) in the phonon band diagram, where $(\sigma_y,\sigma_x)$ labels the propagating mode vector symmetry with respect to reflections about the $xz$ and $yz$ planes respectively. All the other bands at the defect mode frequency are protected by an additional $z=0$ mirror symmetry which is approximately preserved in our fabricated structures. In the case of the optics, we see that the bands accessible via this symmetry breaking are very close to the light line. As described previously, the optical mode is already designed to have  little energy at longitudinal momenta near the light line, and so as long as the perturbation does not introduce additional momentum components in the $x$ direction, the coupling to these bands should be minimal. We can achieve this by inducing the offset uniformly on every hole along the beam. 

\begin{figure}
\includegraphics[width=0.45\textwidth]{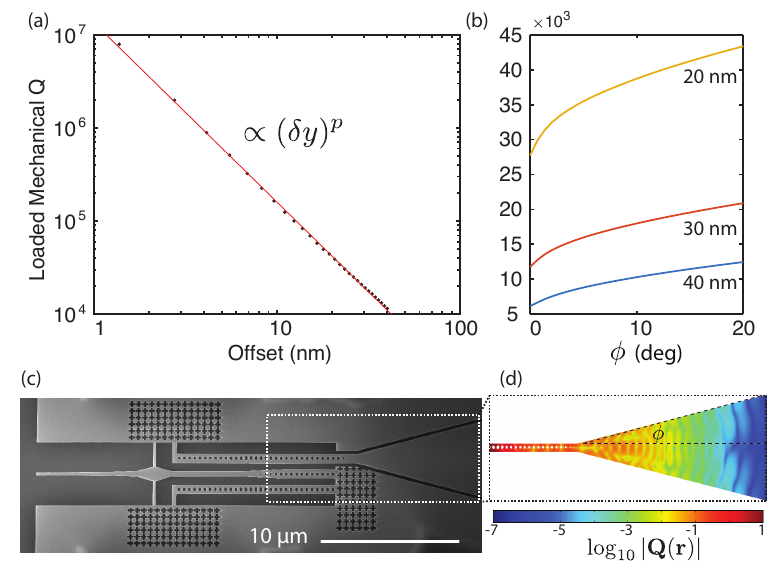}
\caption{\label{sims}System simulation and device fabrication. (a) Nearly inverse square scaling law of the $Q$ with the strength of the offset perturbation. The horn angle is $\phi = 15\degree$ for this simulation. The power law fit gives $p=-1.9$.  (b) Tuning curves showing how the mechanical $Q$ varies with horn taper angle for four different offsets, labeled below the solid lines. $\phi=0\degree$ corresponds to the beam extended infinitely into a perfectly matched layer. (c) A scanning electron microscope image of the experimentally realized device. The horn extends to the right for $15~\micro\meter$. Scalebar inset is $10~\micro\meter$. (d) The mechanical displacement field of a nanobeam coupled to a horn.}
\end{figure}

Finite element simulations in COMSOL are used to verify this type of design. We simulate the optical and mechanical $Q$s of the defect states for values of the offset parameter up to $40~\textrm{nm}$. As expected, the optical $Q$ is not affected with the radiation-limited part staying above $10^7$ for all simulated offsets. The mechanical quality factor is calculated by incorporating a perfectly matched layer (PML) inside a long waveguide connected to the cavity. In this way, the leakage rate can be obtained by determining the complex eigenvalues in finite element analysis. The obtained $Q$s are plotted in Fig.~\ref{sims}a and captures the leakage of phonons into the waveguide. We see a reduction in quality factor that scales approximately as $\delta y^2$. This is expected from a Fermi's golden rule analysis since the coupling matrix element between localized and propagating modes is proportional to $\delta y$.

In a real structure, the waveguide connected to the phononic crystal cavity must be terminated at some point into the silicon slab. To avoid reflections, we design a horn that expands the modes of the beam into surface waves in the silicon slab. We use a linear horn parameterized by an opening angle $\phi$, where $\phi = 0\degree$ corresponds to no change in the width of the beam. The length of the horn in our experiment is $15~\mu \text{m}$, chosen to avoid overlap between neighboring devices. Assuming that the acoustic field is not reflected from the horn output back into the phononic crystal, we can calculate the effect of the horn on the leakage rate for different values of $\phi$. These are plotted in Figure~\ref{sims}b showing that the leakage rate is reduced significantly for large $\phi$. Note that the simulations shown in Fig.~\ref{sims} are done for a symmetric structure and do not take into account the reflection off the boundary of the nanobeam opposite to the horn -- interference  effects can cause non-monotonic changes in loss rate with $\phi$, though for the case studied here they have very little effect on the loading. We choose $\phi=15\degree$ for the fabricated system shown in Fig.~\ref{sims}c.

\section{Experiments}\label{s:experiment}

Our devices are realized on a silicon-on-insulator (SOI) die with a $220~\nano\meter$ device layer on $3~\micro\meter$ buried oxide. Patterns are defined using electron-beam-lithography (JEOL JBX 6300 system using CSAR-62 resist with a $320~\micro\coulomb\per\centi\meter\squared$ dose) followed by plasma etching (Lampoly 9400 TCP Cl$_{2}$/HBr) and release steps to suspend the etched structures. The undercut is performed with a piranha/HF process, followed by a 1:10 HF:H$_{2}$O surface preparation step as described in Ref.~\cite{Hill2013}. The fabricated devices are transferred to a nitrogen flow box to prevent oxidation, before being loaded into the cryostat vacuum chamber. An SEM of the experimentally realized structure is shown in Fig. \ref{sims}c. The data presented in this work is from two separate chips fabricated within the same month, and two separate cool-downs.


\begin{figure}
\includegraphics[width=0.4\textwidth] {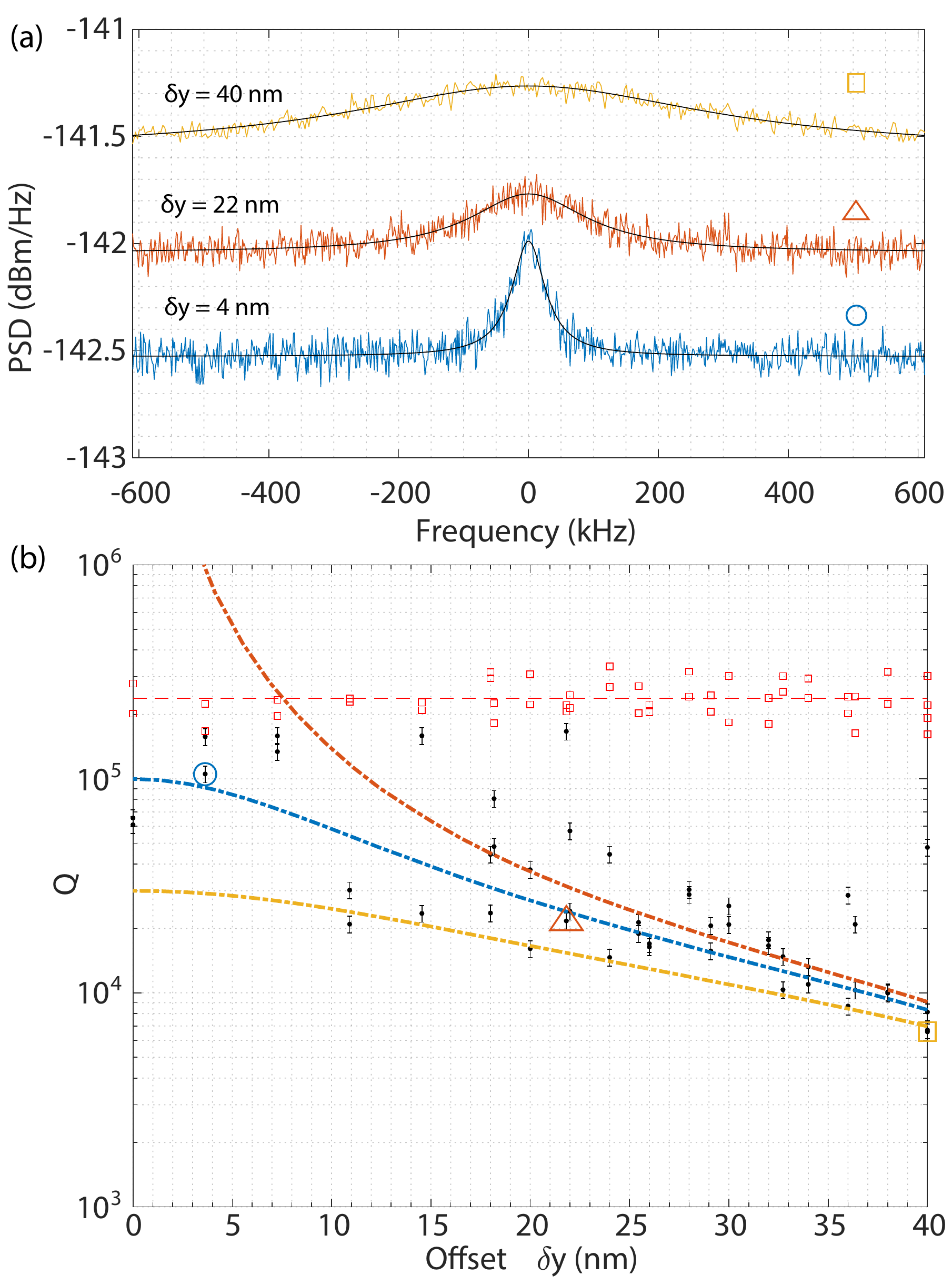}
\caption{\label{expdata} Experimental results. (a) Power spectral density of output fluctuations ($\Delta = \omega_{\textrm{m}})$ showing a thermally driven mechanical resonance at three different offsets. The noise floors are shifted for clarity of comparison. (b) The black data points are mechanical quality factors from the combined dataset, as measured at cryogenic temperatures (sample mount $T = 11~\kelvin$). Error bars reflect a $9\%$ random error, estimated from the standard deviation of repeated measurements on a subset of devices. The red squares show loaded optical quality factors from the combined dataset, with the red dashed line denoting the average loaded optical $Q$. The three dotted dashed lines from bottom to top (yellow, blue, red) are different theory curves from finite element simulations, taking into account the fixed boundary at the nanobeam end opposite the horn, and saturated at intrinsic $Q$s of $3 \cdot 10^{4}$, $10^{5}$ and $10^{7}$ respectively.}
\end{figure}

Our setup consists of a silicon optomechanical crystal optically addressed in a closed-cycle Montana Instruments cryostat operating at $4$ \kelvin. With the driving laser on, the  linearized optomechanical Hamiltonian describing the system in the laser frame is given by
\begin{equation}
\op{H}{}/\hbar = \Delta \aopd \aop + \omega_{\textrm{m}} \bopd \bop + G (\aopd + \aop)(\bopd + \bop)
\end{equation}
where $\aop$ and $\bop$ are cavity sideband photon and phonon annihilation operators, $\Delta = \omega_{\textrm{c}} - \omega_{\textrm{l}}$ is the detuning between the laser and cavity, $\omega_{\textrm{m}}$ is the mechanical frequency, and $G = g_{0}\sqrt{n_{\textrm{c}}}$ is the driven optomechanical coupling rate for a mean intracavity photon number $n_{\textrm{c}}$ and single photon optomechanical coupling $g_{0}$. The fabricated mechanical resonators have a quality factor $Q = \omega_\text{m}/\gamma$, where the linewidth $\gamma = \gamma_\text{in} + \gamma_\text{loading}$, is composed of the $\gamma_\text{in}$, the intrinsic material-limited mechanical loss rate, and $\gamma_\text{loading}$, the loss rate induced by the offset perturbation. Our experiments measure $\gamma$ and its dependence on the offset perturbation to infer how $\gamma_\text{loading}$ is being modified.

Laser light in the 1520-1570 nm band is generated by a tunable external cavity diode laser (NewFocus TLB 6700). We address individual devices with a lensed fiber for optical access with a single pass coupling efficiency up to $\eta  \approx 0.45$. The  reflected light is modulated by the mechanical motion, and can be detected using a high speed photo-detector. The system is operated in the resolved sideband regime, where $\omega_{\textrm{m}} > \kappa$. The resulting power spectral densities for the mechanical motion taken on the red ($\Delta = \omega_{\textrm{m}}$) and blue side ($\Delta = - \omega_{\textrm{m}}$) are fit to obtain mechanical linewidths which will contain contributions due to optomechanical back-action ($\gamma_{\textrm{OM}} = 4G^{2}/\kappa$). On the red (blue) side we have $\gamma_{\textrm{r}} = \gamma + \gamma_{\textrm{OM}}$ and $\gamma_{\textrm{b}} = \gamma - \gamma_{\textrm{OM}}$. At a constant input power ($P\approx 11~\micro\watt$), we then obtain the intrinsic linewidth from: $\gamma = (\gamma_{\textrm{r}} + \gamma_{\textrm{b}})/2$. The results from $48$ devices, showing the mechanical $Q = \omega_{\textrm{m}}/\gamma$ as a function of the offset perturbation $\delta y$ are shown in Fig.~\ref{expdata}b. Separately, we measure the intrinsic mechanical $Q$ of control devices with no offset perturbation and no horn attached. We find, averaging over 11 such devices, that $Q_{\textrm{control}} \approx 1.3 \cdot 10^{5}$, in agreement with previous studies~\cite{Chan2011}. At $40~\nano\meter$ offset we observe a mechanically loaded $Q_{\textrm{min}} = 6.5\cdot 10^{3}$, implying a reduction in phonon lifetime exceeding one decade, with respect to the control devices. Taking $\omega_{\text{m}}/2\pi \approx 4.4~\giga\hertz$, this gives $\gamma/2\pi = 677~\kilo\hertz$, and a phonon loading rate $\gamma_{\textrm{loading}}/2\pi = 643~\kilo\hertz $ showing the phononic resonance is strongly coupled to the out-going mechanical radiation, $\gamma_{\textrm{loading}} \gg \gamma_{\textrm{in}}$. By contrast, the optical $Q$ in the same set of devices, plotted with the red squares, shows no correlation with the offset perturbation. 
We extract the single photon optomechanical coupling rates by measuring the dependence of the mechanical linewidth versus laser power. For a $40~\nano\meter$ offset device we measure $g_{0}/2\pi = 641~\textrm{kHz}$, compared to an adjacent control device (no-offset) where $g_{0}/2\pi = 710~\textrm{kHz}$. We attribute the observed $10\%$ reduction in $g_{0}$ for structures with $40\nano\meter$ offsets to the redistribution of mechanical energy shown in Fig.~\ref{bands_concept}d. This is in agreement with simulations which predict an $8\%$ reduction in $g_{0}$.

\section{Discussion and Conclusions} \label{s:discussion}

\begin{figure}
\includegraphics[width=0.4\textwidth]{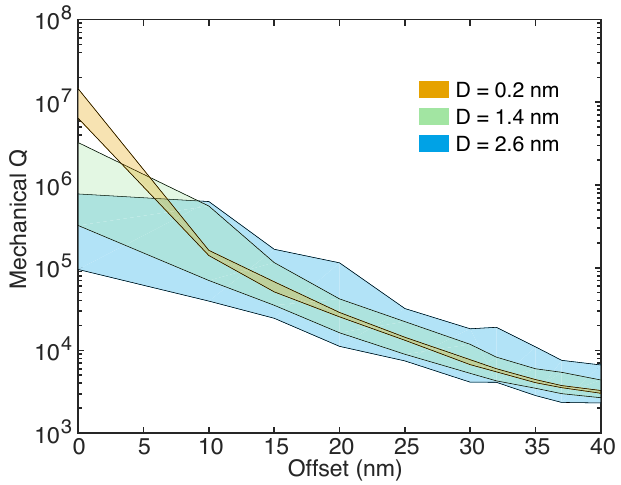}
\caption{\label{sims_disorder} Results of repeated finite element simulations showing how disorder (of standard deviation $D$) affects the mechanical $Q$. The shaded regions are where the middle 68 percent of simulated $Q$s can be found.} 
\end{figure}
In the mechanical $Q$-factor data plotted in Fig.~\ref{expdata}b, we find that larger offsets lead to smaller $Q$ as expected. However, there is significant variation in the mechanical $Q$ between devices with the same offset. For example, for offsets between 20 nm and 25 nm, $Q$s ranging from $2\times 10^3$ to $2\times 10^4$  are experimentally measured. This large variation in the mechanical quality factor can be attributed to an interplay between fabrication disorder and engineered phonon leakage. Unlike photons, for which disorder can cause scattering into the many channels of free space radiation~\cite{Hagino2009}, the number of channels that phonons in this low-dimensional confined geometry can scatter into is very limited. The primary loss channel happens to also be the channel into which we have induced leakage by generating the offset perturbation. This means that fabrication imperfections can cause scattering of phonons into this channel in a way that can interfere both constructively and destructively with the engineered leakage. As a result, in contrast to what is typically the case for optical resonators, the mechanical $Q$ can both decrease \emph{or increase} due to fabrication imperfections. 

To verify this hypothesis, we perform a statistical study by simulation of $>2\times 10^3$ instances of disordered optomechanical crystal cavities (Fig.~\ref{sims_disorder}). Each nanoresonator is simulated with artificially induced disorder. The dimensions of each hole with index $k$ are modified to be $h_{x,k}+\xi_{x,k}$ and $h_{y,k}+\xi_{y,k}$, while the offset of each hole in the $y$ direction is set to be $\delta y_{k}+\xi_{\delta y,k}$. The random variables $\xi_{x,k},~\xi_{y,k}$ and $\xi_{\delta y,k}$ are all normally distributed with standard deviation $D$. At each static offset $\delta y$, multiple simulations are done with varying disorder leading to a number of simulated mechanical $Q$s. These are plotted in Fig.~\ref{sims_disorder} where the shaded regions represent the area where approximately middle $68\%$ of the simulated $Q$s lie. As expected, when the engineered offset is small ($\delta y=0$), disorder causes a reduction in the quality factors. At larger static offsets however, we see that disorder can cause an increase or decrease of the mechanical $Q$ with nearly equal probability.

In conclusion, we have demonstrated that  it is possible to ``open up'' an optomechanical cavity to phonons without affecting the optical properties of the system. The experimentally demonstrated technique will enable engineering phonon networks where highly-coherent mechanical systems are connected to each other across a chip. These experiments also highlight the importance of fabrication disorder in highly confined low dimensional phononic nanostructures.

\textbf{Acknowledgements} This work was supported by NSF ECCS-1509107, ONR MURI QOMAND, and start-up funds from Stanford University. ASN is supported by the Terman and Hellman Fellowships. RNP is supported by the NSF Graduate Research Fellowships Program. WJ was supported by the Tsinghua University undergraduate research program. RNP thanks Rapha{\"e}l van Laer for comments on the manuscript, Dodd Gray for lending an IR camera for the experiment, and Patricio Arrangoiz-Arriola for experimental assistance. Device fabrication was performed at the Stanford Nano Shared Facilities (SNSF) and the Stanford Nanofabrication Facility (SNF). 

%

\end{document}